\documentclass[12pt,letterpaper]{article}

\usepackage{amsmath,amsfonts,amssymb,graphicx,subfigure}
\begin{document}


\begin{flushright} {\footnotesize MIT-CTP-3689\\ HUTP-05/A0044}  \end{flushright}
\vspace{5mm} \vspace{0.5cm}
\begin{center}

\def\thefootnote{\fnsymbol{footnote}}

{\Large \bf The Minimal Model for Dark Matter and Unification} \\[1cm]
{\large Rakhi Mahbubani$^a$, Leonardo Senatore$^{b,}$\footnote{This work is supported in part
by funds provided by the U.S. Department of Energy (D.O.E) under
cooperative research agreement DF-FC02-94ER40818}}
\\[0.5cm]

{\small
\textit{$^a$ Jefferson Physical Laboratory,\\ Harvard University, Cambridge MA 02138}}\\ \vspace{0.6 cm}
{\small
\textit{$^b$ Center for Theoretical Physics, \\
Massachusetts Institute of Technology, Cambridge, MA 02139}}\\

\end{center}
\vspace{.2cm}

 \vspace{.8cm}

\hrule \vspace{0.3cm}
{\small  \noindent \textbf{Abstract} \\[0.3cm]
\noindent Gauge coupling unification and the success of TeV-scale weakly 
interacting dark matter are usually
taken as evidence of low energy
supersymmetry (SUSY).  However, if we assume that the tuning of the higgs
can be explained in some {\it unnatural} way,
from environmental considerations for example, SUSY is no longer a necessary
component of any Beyond the Standard Model theory. In this paper
we study the minimal model with a dark matter candidate and gauge
coupling unification.  This consists of the SM plus fermions
with the quantum numbers of SUSY higgsinos,
and a singlet. It predicts thermal dark matter with a mass that
can range from 100 GeV to around 2 TeV and generically gives rise to
an electric dipole moment (EDM) that
is just beyond current experimental limits,
with a large portion of its allowed parameter space
accessible to next generation EDM and direct detection
experiments.  We study precision unification in this model by
embedding it in a 5-D orbifold GUT
where certain large threshold corrections are calculable, achieving gauge coupling
and $b$-$\tau$ unification, and predicting a rate of proton decay
just beyond current limits.

\vspace{0.5cm}  \hrule

\def\thefootnote{\arabic{footnote}}
\setcounter{footnote}{0}


\section{Introduction\label{int}}

Over the last few decades the search for physics beyond the
Standard Model (SM) has largely been driven by the
principle of naturalness, according to which the parameters of a low
energy effective field theory like the SM should not be much
smaller than the contributions that come from running them up to the
cutoff.  This principle can be used to constrain the
couplings of the effective theory with positive mass dimension,
which have a strong dependence on
UV physics.  Requiring no fine tuning between
bare parameters and the corrections they receive from
renormalization means that the theory must have
a low cutoff.  New
physics can enter at this scale to literally cut off the
high-energy contributions from renormalization.

In the specific case of the SM the effective lagrangian
contains two relevant parameters:  the higgs mass and the cosmological
constant (c.c), both of which
give rise to problems concerning the interpretation of the low
energy theory.  Any discussion of large discrepancies
between expectation and observation must begin with what is known as
the c.c. problem.  This relates to our failure to find a well-motivated
dynamical explanation for the factor of $10^{120}$ between the observed c.c
and the naive contribution to it from renormalization which is proportional
to $\Lambda^4$,
where $\Lambda$ is the cutoff of the theory, usually taken to be equal to the Planck scale.
Until very recently there was still
hope in the high energy physics community that the c.c.
might be set equal to zero by
some mysterious symmetry of quantum gravity.
This possibility has become increasingly unlikely with time since
the observation that our universe is accelerating
strongly suggests the presence of a non-zero
cosmological constant \cite{Perlmutter:1998np,Spergel:2003cb}.

A less extreme example is the hierarchy between the higgs mass
and the GUT scale which can be explained by SUSY breaking
at around a TeV.  Unfortunately the failure of indirect searches
to find light SUSY partners has brought this possibility into question, since
it implies the presence of some small fine-tuning in the SUSY sector.  This `little
hierarchy' problem \cite{Barbieri:1999tm,Giusti:1998gz} raises some
doubts about the plausibility of low energy SUSY as an explanation
for the smallness of the higgs mass.

Both these problems can be understood from a different
perspective:  the fact that the c.c. and the higgs mass are relevant
parameters means that they dominate low energy physics, allowing them
to determine very gross properties of the effective theory.  We might therefore
be able to put limits on them by requiring that this theory
satisfy the environmental conditions necessary
for the universe not to be empty.  This
approach was first used by Weinberg \cite{Weinberg:1987dv} to
deduce an upper bound on the cosmological
constant from structure formation, and was later employed to solve the
hierarchy problem in an analogous way by invoking the atomic
principle \cite{Agrawal:1997gf}.

Potential motivation for this class of argument can be found
in the string theory landscape.  At low energies some regions of the landscape
can be thought of as a field theory with many vacua, each having
different physical properties.  It is possible to imagine that all
these vacua might have been equally populated in the early universe,
but observers can evolve only in the few where the low energy
conditions are conducive to life.  The number of vacua with this
property can be such a small proportion of the total as to dwarf even
the tuning involved in the c.c. problem; resolving the
hierarchy problem similarly needs no further assumptions.
This mechanism for dealing with both issues simultaneously
by scanning all relevant parameters of the low energy theory
within a landscape was recently proposed in \cite{Arkani-Hamed:2004fb,Arkani-Hamed:2005yv}.

From this point of view there seems to be no fundamental inconsistency with
having the SM be the complete theory of our world up to the Planck scale;
nevertheless this scenario presents various problems.  Firstly there is
increasing evidence for dark matter (DM) in the
universe, and current cosmological observations fit well with the
presence of a stable weakly interacting particle at around the TeV scale.
The SM contains no such particle.  Secondly, from a more
aesthetic viewpoint gauge couplings do not quite unify at high energies
in the SM alone; adding weakly interacting particles changes the
running so unification works better.  A well-motivated example of a model that does this is 
Split Supersymmetry \cite{Arkani-Hamed:2004fb}, which is however not the simplest possible theory of this type.  In light of this we study the minimal model
with a finely-tuned
higgs and a good thermal dark matter candidate proposed in \cite{Arkani-Hamed:2005yv}, which also allows for gauge coupling
unification.  Although a systematic analysis of the complete set of such models was carried out in
 \cite{Giudice:2004tc}, the simplest one we study here was missed because the authors
did not consider the possibility of having large UV threshold corrections that
fix unification, as well as a GUT mechanism suppressing proton decay.

Adding just two `higgsino'
doublets\footnote{Here `higgsino' is just a mnemonic
for their quantum numbers, as these particles have nothing to do
with the SUSY partners of the higgs.} to the SM improves unification
significantly.  This model is highly constrained since it contains only one new parameter,
a Dirac mass term for the doublets (`$\mu$'), the neutral components of
which make ideal DM candidates for 990 GeV$\lesssim\mu\lesssim$
1150 GeV (see \cite{Giudice:2004tc} for details).  However a model with pure higgsino dark matter
is excluded by direct detection experiments since the degenerate neutralinos have unsuppressed
vector-like couplings to the $Z$ boson, giving rise to a spin-independent direct detection cross-section that is 2-3 orders of magnitude above current limits\footnote{A model obtained adding a single higgsino doublet, although more minimal, is anomalous and
hence is not considered here.} \cite{Goodman:1984dc,Akerib:2005kh}.
To circumvent this problem, it suffices to include a singlet (`bino') at
some relatively high energy ($\lesssim 10^9$ GeV), with yukawa couplings with the higgsinos and higgs, to lift the mass degeneracy between the
`LSP' and `NLSP'\footnote{From here on we will refer to these particles
and couplings by their SUSY equivalents {\bf without} the quotation marks for simplicity.} by order 100 keV \cite{Smith:2001hy}, as explained in Appendix \ref{app: neutralino mass matrix}.  The instability of such a large mass splitting between the higgsinos and bino to radiative corrections, which tend to make the higgsinos as heavy as the bino, leads us to consider these masses to be separated by at most two orders of magnitude, which is technically natural.  We will see
that the yukawa interactions allow the DM candidate to be as heavy as 2.2 TeV. 
There is also a single reparametrization invariant CP
violating phase which gives rise to a two-loop
contribution to the electron EDM that is well within the reach of next-generation experiments.

Our paper is organized as follows: in Section \ref{sec:model} we briefly introduce the model, in Section \ref{sec:DM} we study the DM relic density in
different regions of our parameter space with a view to constraining these parameters;
we look more closely at the experimental implications of this model in the context of
 dark matter direct detection and EDM experiments in Sections \ref{sec: direct detection} and \ref{sec:EDM}.
Next we study gauge coupling unification at two loops.  We find that this
is consistent modulo unknown UV threshold corrections, however the unification scale is too low
to embed this model in a simple 4D GUT.  This is not necessarily a disadvantage since 4D GUTs have
problems of their own, in splitting the higgs doublet and triplet for example.  A
particularly appealing way to solve all these problems is by embedding our model in a 5D orbifold GUT, in which we can calculate all large threshold corrections and achieve unification.  We also find
a particular model with $b$-$\tau$ unification and a proton lifetime just above current bounds.  We
conclude in Section \ref{sec:conclusion}.

\section{The Model\label{sec:model}}

As mentioned above, the model we study consists of the SM with the addition of
two fermion doublets with the quantum
numbers of SUSY higgsinos, plus a singlet bino, with the following renormalizable interaction terms:
\begin{equation}
\mu\Psi_u \Psi_d+\frac{1}{2}M_1 \Psi_s \Psi_s+\lambda_u\Psi_u h \Psi_s+\lambda_d \Psi_d h^\dag \Psi_s
\end{equation}
\noindent where $\Psi_s$ is the bino, $\Psi_{u,d}$ are the higgsinos, $h$ is the finely-tuned higgs.

We forbid all other renormalizable couplings to SM fields by imposing a parity
symmetry under which our additional particles are odd whereas all SM fields are
even.  As in SUSY conservation of this parity symmetry implies that our LSP is
stable.  

The size of the yukawa couplings between the new fermions and the higgs are limited by
requiring perturbativity to the cutoff.  For equal yukawas this constrains
$\lambda_u(M_Z)=\lambda_d(M_Z)\leq 0.88$, while if we take one of the couplings to be small,
say $\lambda_d(M_Z)=0.1$ then $\lambda_u(M_Z)$ can be as large as 1.38.

The above couplings allow for the CP violating phase $\theta={\rm Arg}(\mu M_1 \lambda^*_u\lambda^*_d)$,
giving 5 free parameters in total.  In spite of
its similarity to the MSSM (and Split SUSY) weak-ino sector, there are a
number of important
differences which have a qualitative effect on the phenomenology of the model,
especially from the perspective of the relic density.
Firstly a bino-like LSP, which usually mixes with the wino, will generically annihilate
less effectively in this model since the wino is absent.
Secondly the new yukawa couplings are free parameters so they can
get much larger than in Split SUSY, where the usual
relation to gauge couplings is imposed at the high SUSY breaking scale.
This will play a crucial
role in the relic density calculation since larger yukawas means greater
mixing in the neutralino sector as well as more efficient annihilation,
especially for the bino which is a gauge singlet.

Our 3$\times$3 neutralino mass matrix is shown below:

\begin{displaymath}
M_N= \left(\begin{array}{ccc}
M_1 & \lambda_u v & \lambda_d v \\
\lambda_u v & 0 &- \mu e^{i\theta} \\
\lambda_d v &-\mu e^{i\theta} & 0
\end{array}
\right) \label{mass_matrix}
\end{displaymath}
for $v=174$ GeV, where we have chosen to put the CP
violating phase in the $\mu$ term.  The chargino is the
charged component of the higgsino with tree level mass
$\mu$.

It is possible to get a feel for the behavior of this matrix by diagonalizing
it perturbatively for small off-diagonal terms, this is done in Appendix
\ref{app: neutralino mass matrix}.

\section{Relic Abundance\label{sec:DM}}

In this section we study the regions of parameter space in which
the DM abundance is in accordance the post-WMAP $2\sigma$
region $0.094<\Omega_{{\rm dm}}h^2<0.129$ \cite{Spergel:2003cb}, where
$\Omega_{{\rm dm}}$ is the fraction of today's critical density in
DM, and $h=0.72\pm 0.05$ is the Hubble constant in units of $100
{\rm \ km/(s\ Mpc)}$. 

As in Split SUSY, the absence of sleptons in our model greatly decreases
the number of decay channels available to the LSP \cite{Pierce:2004mk,Senatore}.  
Also similar to Split SUSY is the fact that our higgs can be heavier than in the
MSSM (in our case the higgs mass is actually a free parameter), hence new decay
channels will be available to it, resulting in a large enhancement of its
width
especially near the $WW$ and $ZZ$ thresholds.  This in turn makes accessible
neutralino annihilation via a resonant higgs, decreasing the relic
density in regions of the parameter space where this channel is
accessible.  For a very bino-like LSP this is easily the dominant
annihilation channel, allowing the bino density to decrease to an
acceptable level.  We use a modified version of the DarkSUSY \cite{DarkSUSY} code for our relic
abundance calculations, explicitly adding the
resonant decay of the heavy higgs to $W$ and $Z$ pairs.

As mentioned in the previous section there are also some differences
between our model and Split SUSY that are relevant to this discussion: the first is that
the Minimal Model contains no wino equivalent (this feature also distinguishes this model from that in \cite{Provenza:2005nq}, which contains a similar dark matter analysis).  The second difference concerns the size of the yukawa
couplings which govern this mixing, as well as the annihilation
cross-section to higgses.  Rather than being tied to the gauge
couplings at the SUSY breaking scale, these couplings are limited only
by the constraint of perturbativity to the cutoff.
This
means that the yukawas can be much larger in our model, helping a bino-like LSP
to both mix more and annihilate more efficiently.  These effects are evident in
our results and will be discussed in more detail below.


We will restrict our study of DM relic abundance and direct detection in this model to the case
with no CP violating phase ($\theta=0,\pi$); we briefly comment
on the general case in Section \ref{sec:EDM}. Our results for different
values of the yukawa couplings are shown in
Figure \ref{dmdata} below, in which we highlight the points in
the $\mu$-$M_1$ plane that give rise to a relic density within the
cosmological bound.  The higgs is relatively heavy ($M_{\rm higgs}=160$ GeV) 
in this plot in order to access processes with 
resonant annihilation through an s-channel higgs.
As we will explain below the only effect this has is
to allow a low mass region for a bino-like LSP with 
$M_1\sim M_{\rm higgs}/2$.

\begin{figure}[ht]
  \centering
    \includegraphics[width=17cm]{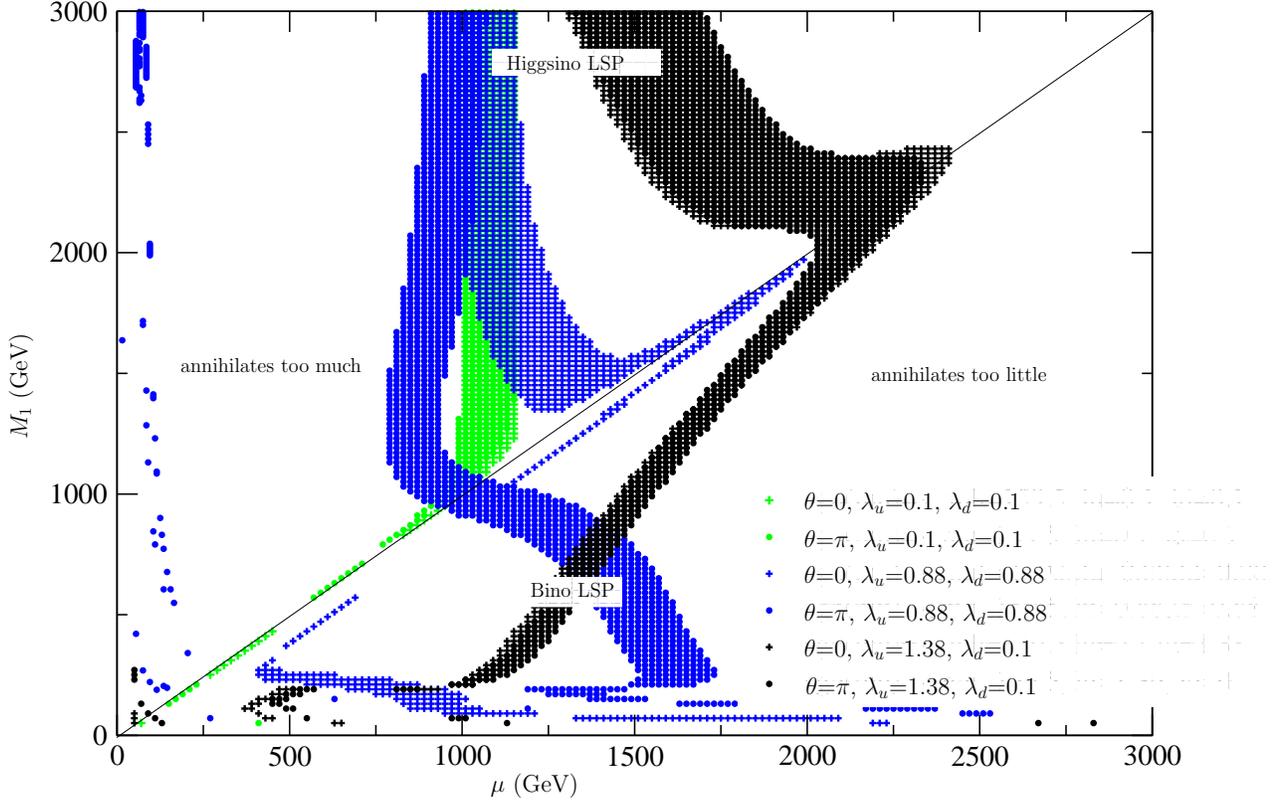}
  \caption{\footnotesize Graph showing regions of parameter space consistent with WMAP.}
  \label{dmdata}
\end{figure}

\begin{figure}[ht]
  \centering
    \subfigure[$\theta$=0]
    {\label{muplus}\includegraphics[width=8 cm]{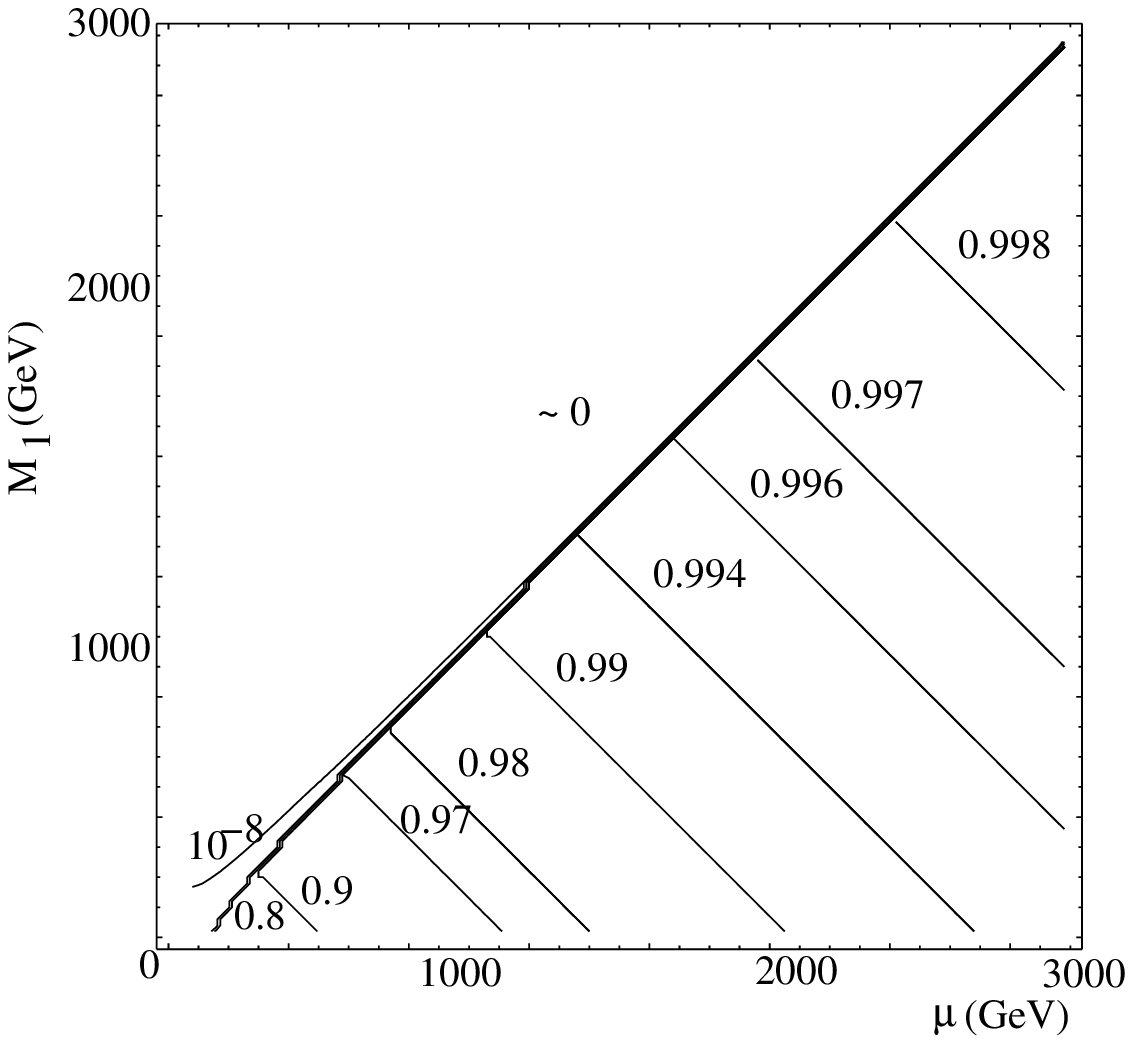}}
    \hspace{0in}
    \subfigure[$\theta$=$\pi$]
    {\label{muminus}\includegraphics[width=8 cm]{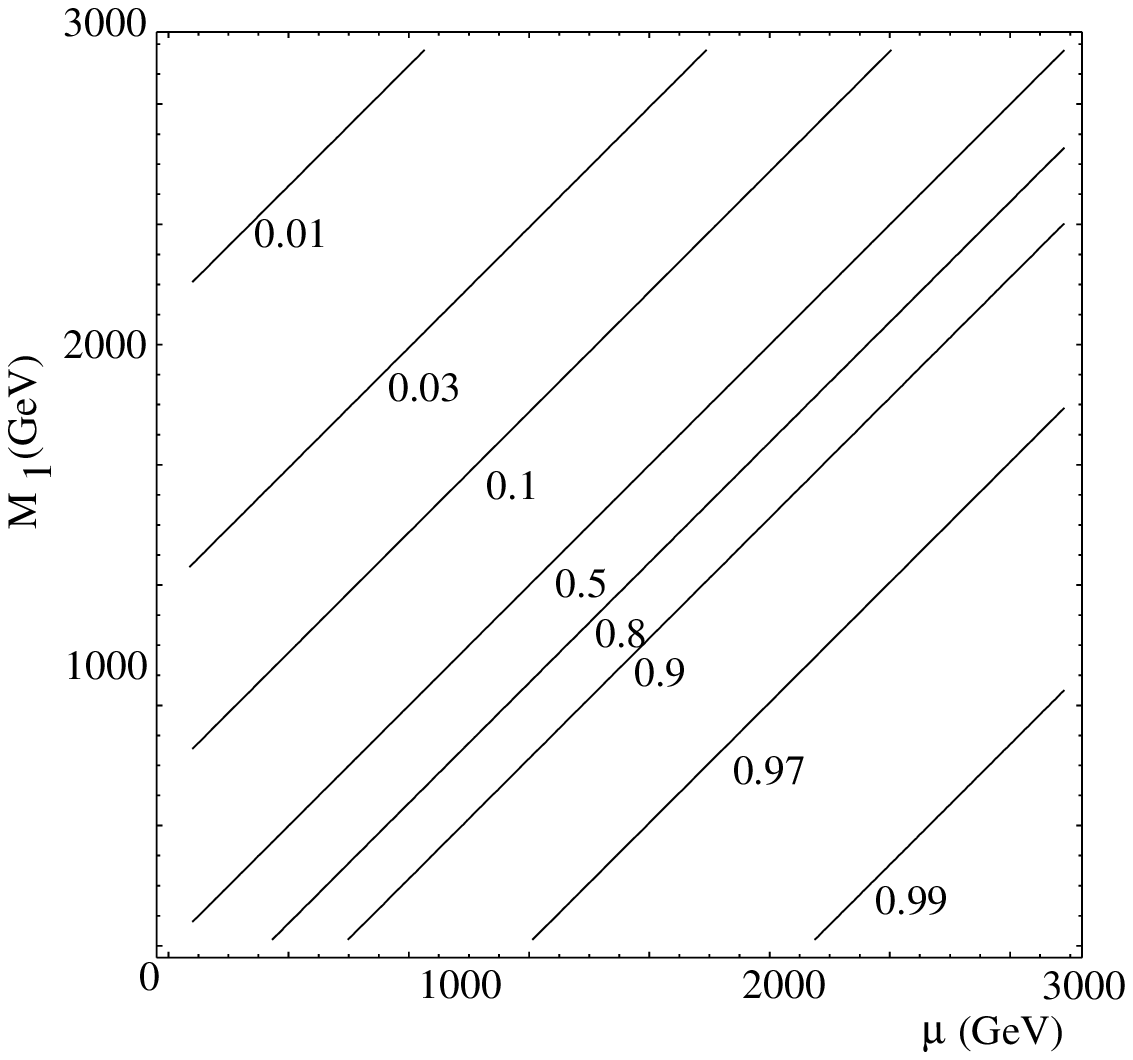}}
  \caption{\footnotesize Gaugino fraction contours for $\lambda_u=\lambda_d=0.88$ and $\theta$=$0$ (left),$\pi$ (right).}
  \label{contourfig}
\end{figure}

Notice that the relic abundance seems to be consistent with a dark
matter mass as large as $2.2$ TeV.  Although a detailed analysis of the LHC signature of this model is not within the scope of this paper, it is clear that a large part of this parameter space will be inaccessible at LHC.  The pure higgsino region for example, will clearly be hard to explore since the higgsinos are heavy and also very degenerate. There is more hope in the bino LSP region for a light enough spectrum.

While analyzing these results we must keep in mind that 
$\Omega_{\rm dm}\sim 10^{-9}{\rm GeV^{-2}}/\langle\sigma\rangle_{\rm eff}$, where $\langle\sigma\rangle_{\rm eff}$
is an effective annihilation cross section for the LSP at the freeze out temperature, which
takes into consideration all coannihilation channels as well as 
the thermal average \cite{Edsjo:1997bg}.  It will be 
useful to approximate this quantity as the cross-section for 
the dominant annihilation channel.  Although rough, this approximation
will
help us build some 
intuition on the behavior of the relic density in different parts of the 
parameter space.  We will not discuss the region close to the origin where the interpretation of the results become more involved due to large mixing and coannihilation.

\subsection{Higgsino Dark Matter}

In order to get a feeling for the structure of Figure \ref{dmdata},
it is useful to begin by looking at the regions in which the physics
is most simple. This can be achieved by diminishing the
the number of annihilation channels that are available to the LSP by taking the limit of small yukawa couplings.  

For $\theta=0$, mixing occurs only on the diagonal $M_1=\mu$ to a very good approximation (see Appendix \ref{app: neutralino mass matrix} and Figure \ref{contourfig}), hence the region above the diagonal corresponds to a pure higgsino LSP with mass $\mu$.  For $\lambda_u=\lambda_d=0.1$ the yukawa interactions are irrelevant and the LSP dominantly annihilates by t-channel neutral (charged) higgsino exchange to $ZZ$ ($WW$) pairs.  Charginos, which have a tree-level mass $\mu$ and are almost degenerate with the LSP, coannihilate with it, decreasing the relic density by a factor of 3.  This fixes the LSP mass to be around $\mu=1$ TeV, giving rise to the wide vertical band that can be seen in the figure; for smaller $\mu$ the LSP over-annihilates, for larger $\mu$ it does not annihilate enough.

Increasing the yukawa couplings increases the importance of t-channel bino exchange to higgs pairs. Notice that taking the limit $M_1\gg\mu$ makes this new interaction irrelevant, therefore the allowed region converges to the one in which only gauge interactions are effective.  Taking this as our starting point, as we approach the diagonal the mass of the bino decreases, causing the t-channel bino exchange process to become less suppressed and increasing the total annihilation cross-section.  This explains the shift to higher masses, which is more pronounced for larger yukawas as expected and peaks along the diagonal where the higgsino and bino are degenerate and the bino mass suppression is minimal.  The increased coannihilation between higgsinos and binos close to the diagonal 
does not play a large part here since both particles have access to a similar t-channel diagram.

Taking $\theta=\pi$ makes little qualitative
difference when either of the yukawas is small compared to $M_1$ or $\mu$,
since in this limit the angle is unphysical
and can be rotated away by a redefinition
of the higgsino fields.  However we can see in Figure \ref{contourfig}
that for large yukawas the region above the diagonal
$M_1=\mu$ changes to a mixed state, rather than being pure higgsino as before.  Starting again with the large $M_1$ limit and decreasing $M_1$ decreases the mass suppression of the t-channel bino exchange diagram like in the $\theta=0$ case, but the LSP also starts to mix more with the bino, an effect that acts in the opposite direction and decreases $\langle\sigma\rangle_{\rm eff}$.  This effect happens to outweigh the former, forcing the LSP to shift to lower masses in order to annihilate enough.  

With $\theta=\pi$ and yukawas large enough, there is an additional allowed region for $\mu<M_W$. In this region the higgsino LSP
is too light to annihilate to on-shell gauge bosons, so the dominant annihilation channels are phase-space suppressed. 
Furthermore if the splitting between the chargino and the LSP is large enough, the 
effect of coannihilation with the chargino into photon and on-shell $W$ is Boltzmann suppressed, substantially decreasing 
the effective cross-section, and giving the right relic 
abundance even with such a light higgsino LSP. Although acceptable from a cosmological standpoint, this
region is excluded by direct searches since it corresponds to a chargino that is too light.

\subsection{Bino Dark Matter}

The region below the diagonal $M_1=\mu$ corresponds to a bino-like
LSP.  Recall that in the absence of yukawa couplings pure binos in
this model do not couple to anything and hence cannot annihilate at
all.  Turning on the yukawas allows them to mix with higgsinos which
have access to gauge annihilation channels.  For
$\lambda_u=\lambda_d=0.1$ this effect is only large enough when $M_1$
and $\mu$ are comparable (in fact when they are equal, the neutralino
states are maximally mixed for arbitrarily small off-diagonal terms),
explaining the stripe near the diagonal in Figure \ref{dmdata}.  Once $\mu$ gets larger than $\sim 1$ TeV even pure higgsinos are too heavy to annihilate efficiently; this
means that mixing is no longer sufficient
to decrease the dark matter relic density to acceptable values and the stripe ends.

Increasing the yukawas beyond a certain value ($\lambda_u=\lambda_d=0.88$, which is slightly larger than their values in Split SUSY, is enough), makes t-channel annihilation to higgses become large enough that a bino LSP does not need to mix at all in order to have the correct annihilation cross-section.  This gives rise to an allowed region which is in the shape of a stripe, where for fixed $M_1$ the correct annihilation cross-section is achieved only for the small range of $\mu$ that gives the right  t-channel suppression.  As $M_1$ increases the stripe converges towards the diagonal in order to compensate for the increase in LSP mass by increasing the cross-section.  Once the
diagonal is reached this channel cannot be enhanced any further, and there is no allowed region for heavier LSPs.  In addition the cross-section
for annihilation through an s-channel resonant higgs, even though
CP suppressed (see Section \ref{sec:EDM} for details), becomes large enough to allow
even LSPs that are very pure bino to annihilate in this way.  The annihilation rate for this process is not very sensitive to the mixing, explaining the apparent horizontal line at $M_1=\frac{1}{2} M_{\rm higgs}\sim 80$ GeV.  This line ends when $\mu$ grows to the point where the mixing is too small.

As in the higgsino case, taking $\theta=\pi$ changes the shape of the contours of constant gaugino fraction and spreads them out in the plane (see Figure \ref{contourfig}), making mixing with higgsinos relevant throughout the region.  For small $M_1$, the allowed region
starts where the mixing term is small enough for the combination of gauge and higgs channels not to cause over-annihilation.  Increasing $M_1$ again makes the region move towards the diagonal, where the increase in LSP mass is countered by increasing the cross-section for the gauge channel from mixing more.

For either yukawa very large ($\lambda_u=1.38, \lambda_d=0.1$), annihilation to higgses via t-channel higgsinos is so efficient that this process alone is sufficient to give bino-like LSPs the correct abundance.  As $M_1$ increases the allowed region again moves towards the diagonal in such a way as to keep the effective cross-section constant by decreasing the higgsino mass suppression, thus compensating for the increase in LSP mass.  As we remarked earlier since $\lambda_d$ is effectively zero in this case, the angle $\theta$ is unphysical and can be rotated away by a redefinition of the higgsino fields.

\section{Direct Detection\label{sec: direct detection}}

Dark matter is also detectable through elastic scatterings off ordinary matter.
The direct detection cross-section for this process
can be divided up into a spin-dependent and a spin-independent part; we will
concentrate on the former since it is usually dominant.  As before we
restrict to $\theta=0$ and $\pi$, we expect the result not to change significantly
for intermediate values.

The spin-independent
interaction takes place through higgs exchange, via the yukawa couplings
which mix higgsinos and binos.  Since the only $\chi^0_1\chi^0_1 h$ term in our
model involves the
product of the gaugino and higgsino fractions, the more mixed our dark matter is
the more visible it will be to direct detection experiments.  This effect can be
seen in Fig \ref{fig: direct detection} below.

\begin{figure}[ht]
  \centering
    \includegraphics[width=17cm]{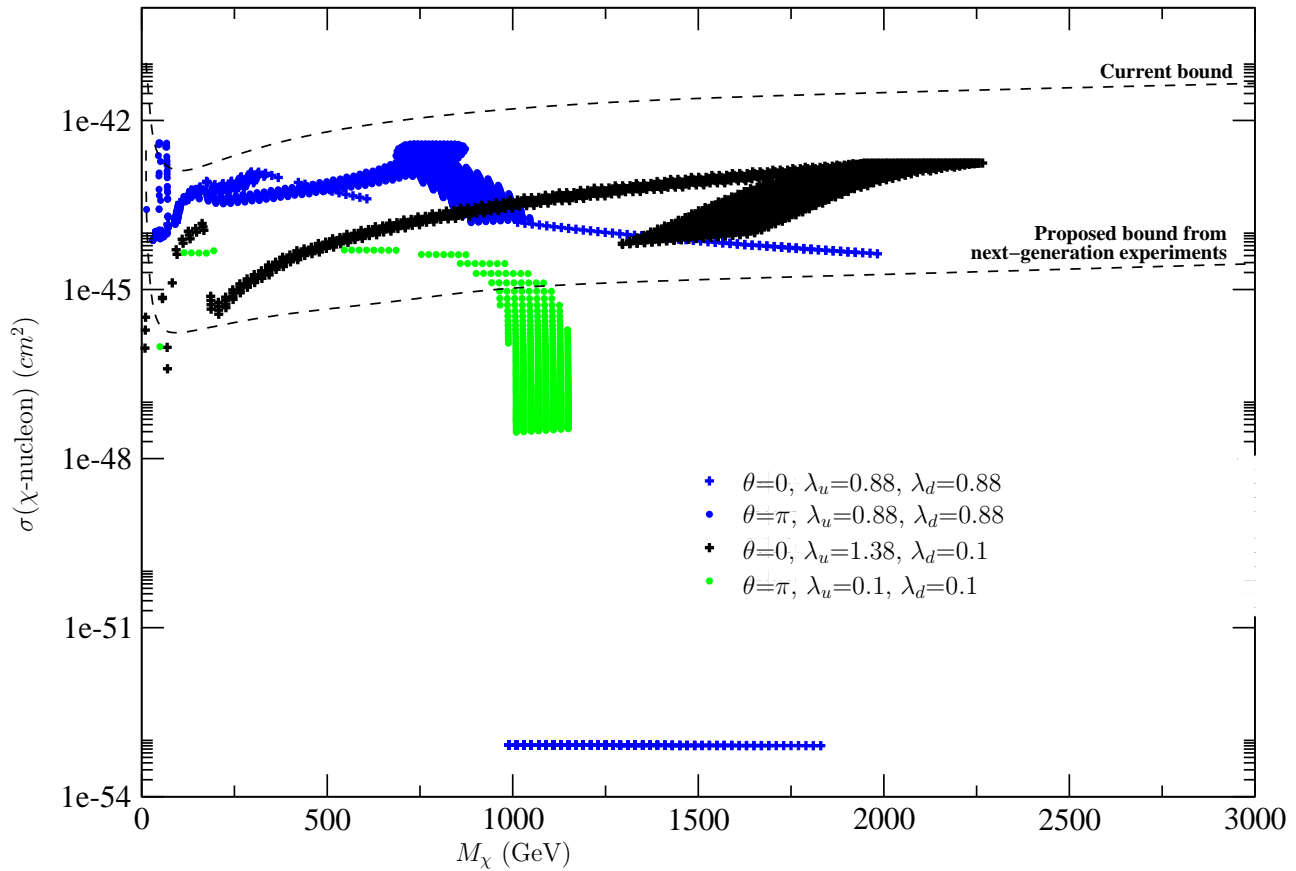}
  \caption{\footnotesize Spin-independent part of dark matter direct detection cross-section. The current bound represents the CDMS limit \cite{Akerib:2005kh},
and, as an indicative value for the proposed bound from next generation experiments, we take the projected sensitivity of SuperCDMS 
phase B \cite{Brink:2005ej}.}
  \label{fig: direct detection}
\end{figure}

\noindent Although it seems like we cannot currently use this measure as a constraint,
the major proportion of our parameter space will be accessible at
next-generation experiments.  Since higgsino LSPs are generally more pure than
bino-type ones, the former will escape detection as
long as there is an order 100 keV splitting between its two neutral components.
This is is necessary in order to avoid the limit from spin-independent direct
detection measurements \cite{Smith:2001hy}.

Also visible in the graph are the interesting discontinuities mentioned in
\cite{Pierce:2004mk}, corresponding to the opening up of new annihilation
channels at $M_{LSP}=1/2 M_{higgs}$ through an s-channel higgs.  We also
notice a similar discontinuity at the top threshold from annihilation
to $t\overline{t}$; this effect becomes more pronounced
as the new yukawa couplings increase.

\section{Electric Dipole Moment\label{sec:EDM}}

Since our model does not contain any sleptons it
induces an electron EDM only at two loops, proportional to $\sin(\theta)$
for $\theta$ as defined above.  This is a two-loop effect, we
therefore expect it to be close to the
experimental bound for ${\cal O}(1)$ $\theta$.
The dominant diagram responsible for the EDM is generated by charginos and neutralinos in a loop
and can be seen in Figure \ref{fig:2loop} below.  This
diagram is also present in Split SUSY where it gives a comparable contribution to the one with
only charginos in the loop \cite{Chang:2005ac,Arkani-Hamed:2004yi}.

\begin{figure}[ht]
   \centering
    \includegraphics[trim = 120 0 0 0]{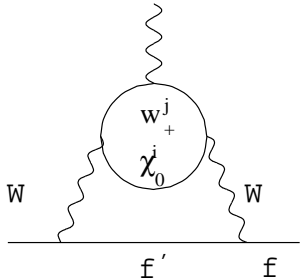}
  \caption{\footnotesize The  2-loop contribution to the EDM of a fermion f.}
  \label{fig:2loop}
\end{figure}

The induced EDM is (see \cite{Chang:2005ac}):

\begin{equation}
\label{eq:WEDM} {d^W_f\over e} =  \pm \frac{\alpha^2 m_f }
{8 \pi^2 s^4_W M_W^2}
 \sum_{i=1}^3  {m_{\chi_i}\mu
\over M_W^2} \hbox{ Im }(O^L_{i} O^{R*}_{i})
 {\cal G}\left( r^0_i, r^\pm \right)
\end{equation}

\noindent where

\begin{eqnarray}
 {\cal G}\left( r^0_i, r^\pm \right)
 &=& \int^\infty_0 dz\int^1_0 { d\gamma \over \gamma} \int^1_0  dy\;
{y\, z\, (y +z/2 )\over (z+y)^3(z+K_{i})}\nonumber\\\nonumber
&=& \int^1_0 { d\gamma \over \gamma} \int^1_0 dy\, y  \left[
{(y-3K_{i})y+2(K_{i}+y)y \over 4 y(K_{i}-y)^2
}+{K_{i}(K_{i}-2y) \over 2(K_{i}-y)^3}\ln\frac{K_{i}}{y}
 \right]
\end{eqnarray}

\noindent and
\begin{eqnarray} &&K_{i}= {r^0_i \over
1-\gamma}+{r^\pm \over \gamma},  \hspace{0.4 cm} r^\pm \equiv
{\mu^2 \over M_W^2}, \hspace{0.4 cm} r^0_i \equiv {m_{\chi_i}^2 \over
M_W^2}, \nonumber\\ \nonumber
&&O^R_{i}=\sqrt{2} N_{2i}^* \exp^{-i\theta}, \hspace{0.4 cm}
O^L_{i}=- N_{3i}\
\end{eqnarray}

\noindent $N^T M_N N=\rm{diag}(m_{\chi_1},m_{\chi_2},m_{\chi_3})$
with real and positive diagonal elements.  The sign on the right-hand side of
equation (\ref{eq:WEDM}) corresponds to the fermion $f$ with weak isospin $\pm
\frac{1}{2}$ and $f'$ is its electroweak partner.

In principle it should be possible to cross-correlate the region of our
parameter space which is consistent with relic abundance
measurements, with that consistent with electron EDM measurements in order to further constrain our parameters.
However since the current release of DarkSUSY does not support CP violating phases and a version including CP violations seems almost ready for public release\footnote{Private communication with one of the authors of DarkSUSY.} we leave an accurate study of the consequences of non-zero CP phase in relic abundance and direct detection calculations for a future work.  We can still draw some interesting conclusions by estimating the effect of non-zero CP phase.
Because there is no reason for these new contributions to be suppressed
with respect to the CP-conserving ones (for $\theta$ of ${\cal{O}}(1)$), we might naively expect their
inclusion to enhance the annihilation cross-section
by around a factor of 2,
increasing the acceptable LSP masses by $\sim\sqrt{2}$ for constant
relic abundance.  This is discussed in greater detail in
\cite{Gondolo:1999gu} (and \cite{Falk:1999mq} for direct detection) in which we see that this observation holds for most of the parameter space.  We must note, however, that in particular small regions of the space the enhancement to the annihilation cross-section and the suppression to the elastic cross section can be much larger, justifying further investigation of this point in future work.  With this
assumption in mind we see in Figure \ref{fig: edm} that although the
majority of our allowed region
is below the current experimental limit of $d_e<1.7\times 10^{-27}e \
{\rm cm}$ at $95\%$ C.L. \cite{edmex}, most of it will be
accessible to next generation EDM experiments.  These propose to
improve the precision of the electron EDM measurement by 4 orders of magnitude
in the next 5 years, and maybe even up to 8 orders of magnitude,
funding permitting \cite{demille,lamo,seme}. We also see in this figure that
CP violation is enhanced on the diagonal where the mixing is largest.
This is as expected
since the yukawas that govern the mixing are necessary for there to be any CP violating phase at all.
For the same reason, decoupling either particle sends the EDM to zero.

\begin{figure}[ht]
  \centering
    \includegraphics[width=17cm]{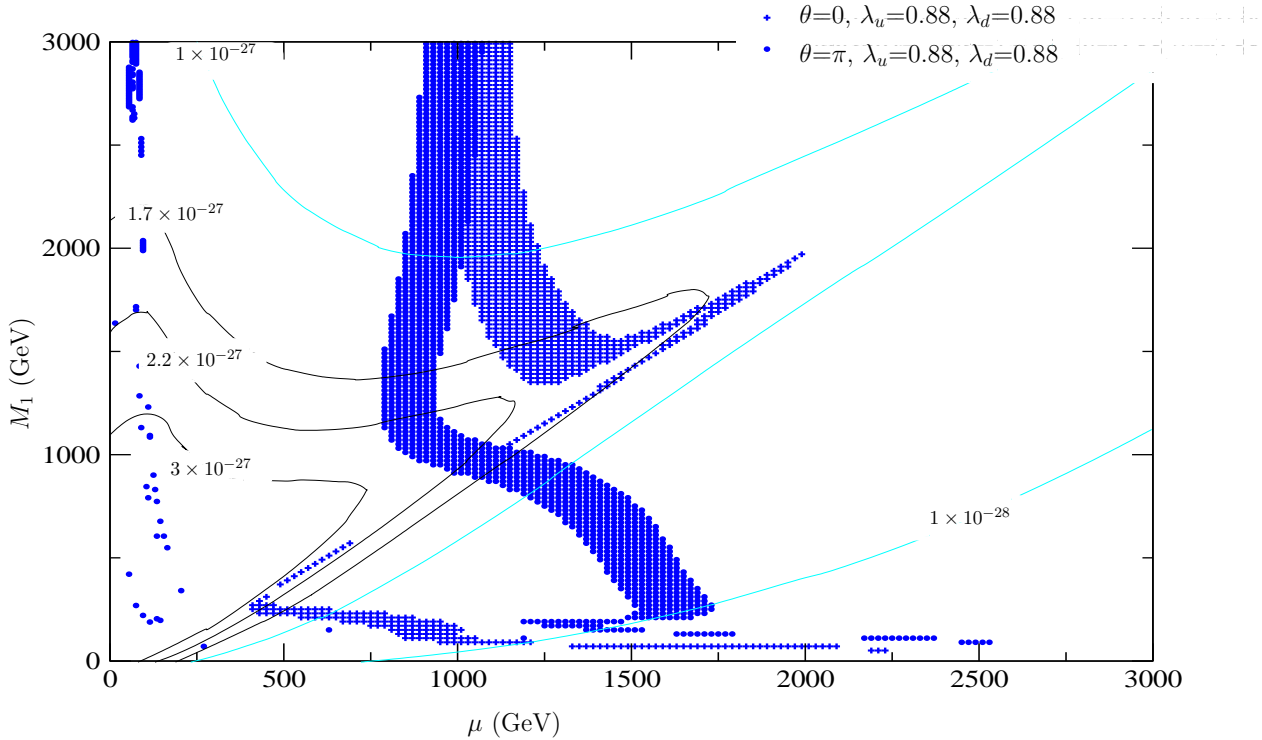}
  \caption{\footnotesize Electron edm contours for $\theta=\pi/2$.  The excluded
region is bounded by the black contours.  Note that CP violation was not included in the relic density calculation, and the dark matter plot is simply intended to indicate the approximate region of interest for dark matter.}
  \label{fig: edm}
\end{figure}

\section{Gauge Coupling Unification \label{sec:unification}}

In this section we study the running of gauge couplings in our
model at two loops.  The addition of higgsinos largely improves
unification as compared to the SM case, but their
effect is still not large enough and the model predicts a value for $\alpha_s(M_z)$
around 9$\sigma$ lower than the experimental value of
$\alpha_s(M_z)=0.119\pm 0.002$ \cite{Eidelman:2004wy}.
Moreover the scale at which the couplings unify is very low,
around $10^{14}$ GeV, making proton decay occur much too
quickly to embed in a simple GUT theory\footnote{It is possible to evade the constraint from proton decay by setting some of the relevant mixing parameters to zero \cite{Dorsner:2004xa}.  However we are not aware of any GUT model in which such an assertion is justified by symmetry arguments.}.
These problems can be avoided by adding the Split SUSY particle content at higher energies, as in \cite{antoniadis}, at the cost of losing minimality; instead we choose to solve this problem by embedding our minimal model in an extra
dimensional theory. This decision is well
motivated: even though normal 4D GUTs have had
some successes,  explaining the quark-lepton mass relations for
example, and charge quantization in the SM, there are many reasons
why these simple theories are not ideal.  In spite of the
fact that the matter content of the SM falls naturally into
representations of $SU(5)$, there are some components that seem
particularly resistant to this.  This is especially true of the higgs sector,
the unification of which gives rise to the doublet-triplet splitting problem.
Even in the matter sector, although $b$-$\tau$ unification works
reasonably well the same cannot be said for unification of the first two generations.
In other words, it seems like gauge couplings want to unify while
the matter content of the SM does not, at least not to the same extent.
This dilemma is easily addressed in an extra dimensional
model with a GUT symmetry in the bulk, broken down to the SM gauge group
on a brane by boundary conditions \cite{Kawamura:1999nj}
since we can now choose where we put fields based on whether
they unify consistently or not.  Unified matter can be placed in the
bulk whereas non-unified matter can be placed on the
GUT-breaking brane.  The low energy theory will then contain the zero modes of the 5D bulk fields
as well as the brane matter.  While solving many of the
problems of standard 4D GUTs these extra dimensional theories
have the drawback of having a large number of discrete choices for
the location of the matter fields, as we shall see later.

We will consider a model with one flat extra dimension compactified on
a circle of radius $R$, with orbifolds $S^1/(Z_2\times Z'_2)$,
whose action is obtained from the identifications

\begin{equation}
Z_2: y\sim 2\pi\ R-y, \ \ \ Z'_2:y\sim\pi R-y,\ \ y\ \epsilon\
[0,2\pi]
\end{equation}

\noindent where $y$ is the coordinate of the fifth dimension. There are two
fixed points under this action, at $(0,\pi R)$ and $(\pi R/2,3\pi
R/2)$, at which are located two branes.  We impose an $SU(5)$
symmetry in the bulk and on the $y=0$ brane; this symmetry is
broken down to the SM $SU(3)\times SU(2)\times U(1)$ on the other
brane by a suitable choice of boundary conditions.  All fields
need to have definite transformation properties under the orbifold
action - we choose the action on the fundamental to be
$\phi\rightarrow \pm P\phi$ and on the adjoint, $\pm [P,A]$, for
projection operators $P_Z=(+,+,+,+,+)$ and $P_{Z'}=(+,+,+,-,-)$.
This gives SM gauge fields and their corresponding KK towers $A^a_\mu$ for $a=\{1,...,12\}$
$(+,+)$ parity; and the towers
$A^{\hat{a}}_\mu$ for $\hat{a}=\{13,...,24\}$ $(+,-)$ parity, achieving the
required symmetry-breaking pattern. By gauge invariance the unphysical fifth component of the gauge field, which is eaten in unitary gauge, gets opposite boundary conditions.\footnote{From an effective field theory
point of view an orbifold is not absolutely necessary, our theory can simply be
thought of as a theory with a compact extra dimension on an
interval, with two branes on the boundaries. Because of the
presence of the boundaries we are free to impose either Dirichlet
or Neumann boundary conditions for the bulk field on each of the
branes breaking the $SU(5)$ to the SM gauge group purely by choice
of boundary conditions and similarly splitting the multiplets
accordingly. Our orbifold projection is therefore nothing more
than a further restriction to the set of all possible choices we
can make.}  We still have the freedom to
choose the location of the matter fields. In this model $SU(5)$-symmetric matter
fields in the bulk will get split by the action of the $Z'$
orbifold: the SM ${\bf \overline{5}}$ for instance will either
contain a massless $d^c$ or a massless $l$, with the other
component only having massive modes.  Matter fields in the
bulk must therefore come in pairs with opposite eigenvalues under the
orbifold projections, so for each SM generation in the bulk we will
need two copies of ${\bf 10} + {\bf \overline{5}}$.  This provides
us with a simple mechanism to forbid proton decay from $X$- and $Y$-exchange
and also to split the color triplet higgs field from the doublet.
To summarize, unification of SM matter fields in complete multiplets of $SU(5)$
cannot be achieved in the bulk but on the $SU(5)$ brane, while matter
on the SM brane is not unified into complete GUT representations.

\subsection{Running and matching}

We run the gauge couplings from the weak scale to the cutoff
$\Lambda$ by treating our model as a succession of effective field
theories (EFTs) characterized by the differing particle content at
different energies.  The influence of the yukawa couplings between
the higgsinos and the singlet on the two-loop running is
negligible, hence it is fine to assume that the singlet is
degenerate with the higgsinos so there is only one threshold from
$M_{\rm top}$ to the compactification scale $1/R$, at which we
will need to match with the full 5D theory.

The $SU(5)$-symmetric bulk gauge coupling $g_5$ can be
matched on to the low energy couplings at the renormalization
scale $M$ via the equation

\begin{equation}\label{matching}
\frac{1}{g^2_i(M)}=\frac{2\pi
R}{g^2_5}+\Delta_i(M)+\lambda_i(M R)
\end{equation}

\noindent The first term on the right represents a tree level contribution from the
5d kinetic term, $\Delta_i$ are similar contributions from
brane-localized kinetic terms and $\lambda_i$ encode radiative
contributions from KK modes.  The latter come from
renormalization of the 4D brane kinetic terms which run
logarithmically as usual.

To understand this in more detail let us consider radiative
corrections to a $U(1)$ gauge coupling in an extra dimension
compactified on a circle with no orbifolds, due to a 5D massless
scalar field \cite{Contino:2001si}.  Since $1/g^2_5$ has mass
dimension 1,  by dimensional analysis we might expect corrections
to it to go like $\Lambda+m\log{\Lambda}$ where $m$ is some mass
parameter in the theory. The linearly divergent term is UV
sensitive and can be reabsorbed into the definition of $g_5$,
whereas the log term cannot exist since there is no mass parameter
in the theory.  Hence the 5D gauge coupling does not run, and
neither does the 4D gauge coupling. This can also be interpreted from a 4D point of view, where the KK partners of the scalar cut off the divergences of the
zero mode.   Since there is no distinction between the
wavefunctions for even (cosine) and odd (sine) KK modes in the absence
of an orbifold, and we know that the sum
of their contributions must cancel the log divergence of the 4D
massless scalar, each
of these must give a contribution equal to $-1/2$ times
that of the 4D massless scalar.

When we impose a $Z_2$ orbifold projection and add two 3-branes at
the orbifold fixed points, the scalar field
must now transform as an eigenstate of this orbifold action and can
either be even ($(+,+)$, with Neuman boudary conditions on the
branes), or odd ($(-,-)$, with Dirichlet boundary conditions on the
branes).  This restricts us to a subset of the original modes and
the cancellation of the log divergence no longer works.  Since this
running can only be due to 4D gauge kinetic terms localized on the branes, where the gauge coupling is dimensionless and can therefore receive logarithmic corrections, 
locality implies that the contribution from a
tower of states on a particular brane can only be due to its
boundary condition on that brane, with the total running equal to
the sum of the contributions on each brane.  In fact, it is only in the vicinity of the brane that imposing a particular boundary condition has any effect.  As argued
above, a $(+,+)$ tower (excluding the zero mode) and a $(-,-)$ tower
must each give a total contribution equal to $-1/2$ times that of
the zero mode, which corresponds to a coefficient of $-1/4$ to the running of each brane-localized kinetic term.  Taking into account the contribution of the zero mode we can
say that a tower of modes with $+$ boundary conditions on a brane
contributes $+1/4$ times the corresponding 4D coefficient, while a
$-$ boundary condition contributes $-1/4$ times the same quantity.
This argument makes it explicit that the orbifold projection can be seen as a prescription on the boundary conditions of the fields in the extra dimension, which only affect the physics near each brane. 

Adding another orbifold projection as we are doing in this case also
allows for towers with $(+,-)$ and $(-,+)$ boundary conditions which, from the above argument, both give a contribution of $\pm 1/4 \mp 1/4=0$.  The contribution of the $(+,+)$ and $(-,-)$ towers clearly remains unchanged.

Explicitly integrating out the KK modes at one loop at the compactification
scale allows us to verify this fact, and also compute the constant parts
of the threshold corrections, which are scheme-dependent.  In $\overline{{\rm
DR}}$ \footnote{We use this renormalization scheme even though our
theory is non-supersymmetric since 4D threshold corrections in this scheme
contain no constant part \cite{Antoniadis:1982vr}.} we obtain
\cite{Contino:2001si}:

\begin{equation}
\lambda_i(M R)=\frac{1}{96\pi^2}\left(\left(b^S_i-21 b^G_i+8
b^F_i\right) F_e(M R)+\left(\tilde{b}^S_i-21\tilde{b}^G_i+8
\tilde{b}^F_i\right) F_0\right)
\end{equation}

\noindent with

\begin{eqnarray}
&&F_e(\mu R)={\cal{I}}-1-\log(\pi)-\log(M R), \ \ \ F_0=-\log(2)\\
\nonumber && {\cal{I}}=\frac{1}{2}\int^{+\infty}_1 dt \
\left(t^{-1}+t^{-1/2}\right)\left(\theta_3(i t)-1\right)\simeq
0.02, \ \ \ \theta_3(i t)=\sum^{+\infty}_{n=-\infty} e^{-\pi t
n^2}
\end{eqnarray}

\noindent where
$b^{S,G,F}_i$ ($\tilde{b}^{S,G,F}_i$) are the Casimirs of the KK modes of
real scalars (not including
goldstone bosons), massive vector bosons and Dirac fermions respectively with even (odd)
masses $2n/R$ ($(2n+1)/R$).
As explained above, the logarithmic part of the above expression
is equal to exactly $-1/2$ times the contribution of the same fields in
4D \cite{Weinberg:1980wa,Hall:1980kf}.  Since the compactification scale $1/R$ will
always be relatively close to the unification scale
$\Lambda$ (so our 5D theory remains perturbative), it will be sufficient for us to use one loop
matching in our two loop analysis as long as the matching is done
at a scale $M$ close to the compactification scale.

As an aside, from equation (\ref{matching}) we can get:

\begin{equation}\label{equ:deltarunning}
\frac{d}{d t}\Delta_i=\frac{b_i-b^{MM}_i}{8\pi^2}
\end{equation}

\noindent where $b_i$ is shorthand for the combination $(b^S_i-21
b^G_i+8 b^F_i)/12$ and $b^{MM}_i$ are the coefficients of the
renormalization group equations below the compactification scale
(see Appendix \ref{app: beta functions} for details). It is clear
from this equation that it is unnatural to require
$\Delta_i(1/R)\ll 1/(8\pi^2)$. The most natural assumption
$\Delta_i(\Lambda)\sim 1/(8\pi^2)$ gives a one-loop contribution comparable
to the tree level term, implying
the presence of some strong dynamics in the brane gauge sector at the scale $\Lambda$. We
know that the 5D gauge theory becomes strong at the scale
$24\pi^3/g^2_5$, so from naive dimensional analysis (NDA) (see for
example \cite{Chacko:1999hg}) we find that it is quite natural for
$\Lambda$ to coincide with the strong coupling scale for the bulk gauge
group. 

Running equation (\ref{equ:deltarunning}) to the compactification scale we obtain

\begin{equation}
\Delta_i(1/R)=\Delta_i(\Lambda)+\frac{b_i-b^{MM}_i}{8\pi^2}\log(\Lambda
R)
\end{equation}

\noindent For $\Lambda R\gg 1$ the unknown bare parameter is negligible
compared to the log-enhanced part, and can be ignored, leaving us with a 
calculable correction.
Using $g_{GUT}=2\pi R/g^2_5$ we expect $\Lambda R\sim 8\pi^2/g^2_{{\rm
GUT}}\sim 100$. Keeping this in mind,
we shall check whether unification is possible in our model with $\Lambda
R$ in the regime where the bare brane gauge coupling is negligible.
To this purpose we will impose the matching equation
(\ref{matching}) at the scale $\Lambda$ {\it assuming}
$\Delta_i(\Lambda)=0$; we will then check whether the value of $\Lambda
R$ found justifies this approximation.

In order to develop some intuition for the direction that these thresholds
go in, we can analyze the one-loop expression (with one-loop thresholds) for the gauge
couplings at $M_z$:

\begin{equation}
\frac{1}{\alpha_i(M_z)}=\frac{4\pi}{g^2_{\rm GUT}}+4\pi
\lambda_i(\Lambda R)+\lambda^{\rm conv}_i(\Lambda
R)+\frac{b_i^{\rm MM}}{2\pi}\log\left(\frac{\Lambda}{M_z}\right)+\frac{\left(b^{\rm
SM}_i-b_i^{\rm MM}\right)}{2\pi}\log\left(\frac{\mu}{M_z}\right)
\end{equation}

\noindent $b^{\rm SM}_i$ are the SM beta function coefficients (see
Appendix \ref{app: beta functions}), $\mu$ is the scale of the
higgsinos and singlet, $\lambda^{\rm
conv}=(-\frac{3}{12\pi},-\frac{2}{12\pi},0)$ are conversion
factors from $\overline{{\rm MS}}$, in
which the low-energy experimental values for the gauge couplings
are defined, to $\overline{{\rm DR}}$ \cite{Langacker:1992rq}.  Taking the linear combination
$(9/14)\alpha_1^{-1}-(23/14)\alpha_2^{-1}+\alpha_3^{-1}$ allows us
to eliminate the $\Lambda$ dependence as well as all
$SU(5)$-symmetric terms, leaving

\begin{equation}
\frac{1}{\alpha_3(M_z)}=-\frac{9/14}{
\alpha_1(M_z)}+\frac{23/14}{\alpha_2(M_z)} +4\pi \lambda(\Lambda R)+
\lambda^{\rm conv}+\frac{\left(b^{\rm
SM}-b^{\rm MM}\right)}{2\pi}\log\left(\frac{\mu}{M_z}\right)
\end{equation}

\noindent where $X=(9/14)X_1-(23/14) X_2+ X_3$ for any quantity $X$.
Recall that the leading threshold correction from the 5D
GUT is proportional to $\log(\Lambda R)$.  The low-energy
value of $\alpha_3$ is therefore changed by

\begin{equation}
\delta\alpha_3(M_z)=\alpha_3(M_z)^2\frac{b}{2\pi}\log(\Lambda R)
\end{equation}

We still have the freedom to choose the positions of the various
matter fields.  In order to determine the best setup
for gauge coupling unification we need to keep in mind two facts:
the first is that adding $SU(5)$ multiplets in the bulk does not have any
effect on $\alpha_3(M_Z)$; and the second is that $b$ contains
only contributions from $(+,+)$ modes (in unitary gauge 
none of our bulk modes have
$(-,-)$ boundary conditions; our $SU(5)$ bulk multiplets are split into $(+,+)$ and $(+,-)$ modes).

As stated at the beginning of this section our 4D prediction for
$\alpha_3(M_z)$ is too low.  Since fermions have a larger effect
on running than scalars, this problem is most efficiently tackled by
splitting up the fermion content of the SM into non-$SU(5)$
symmetric parts in order to make $b$ as positive as possible.
Examining the particular linear combination that eliminated the
dependence on $\Lambda$ at one loop we find that one or more
$SU(5)$-incomplete colored multiplets are needed in the bulk, or
equivalently the weakly-interacting part of the same multiplet has
to be on one of the branes.  Since matter in the bulk is naturally
split by the orbifold projections, this just involves separating
the pair of multiplets whose zero modes make up one SM family.

With this in mind we find that for fixed $\Lambda R$ and $\mu$,
since separating different numbers of SM generations allows us to vary
the low energy value of $\alpha_3$ anywhere from its experimental
value to several $\sigma$s off, gauge coupling unification really
does work in this model for some fraction of all available configurations.
Although this may seem a little unsatisfactory from the
point of view of predictivity, the situation can be somewhat
ameliorated by further refining our requirements.  For example, we can go some way towards explaining the hierarchy between the SM fermion masses by placing the first generation in the bulk, the second generation split between the bulk and a brane and the third generation entirely on a brane.  This way, in addition to breaking the approximate flavor symmetry in the fermion sector we also obtain helpful factors of order $1/\sqrt{\Lambda R}$ between the masses of the different generations.  
The location
of the higgs does not have a very large effect on
unification, the simplest choice would be to put it, as well as
the higgsinos and singlet, on the $SU(5)$-breaking brane, where there
is no need to introduce corresponding color triplet fields.  This also helps to explain the hierarchy.
\begin{figure}[ht]
  \centering
    \subfigure[Higgs on the broken brane]
    {\label{smallestmodel}\includegraphics[width=7 cm]{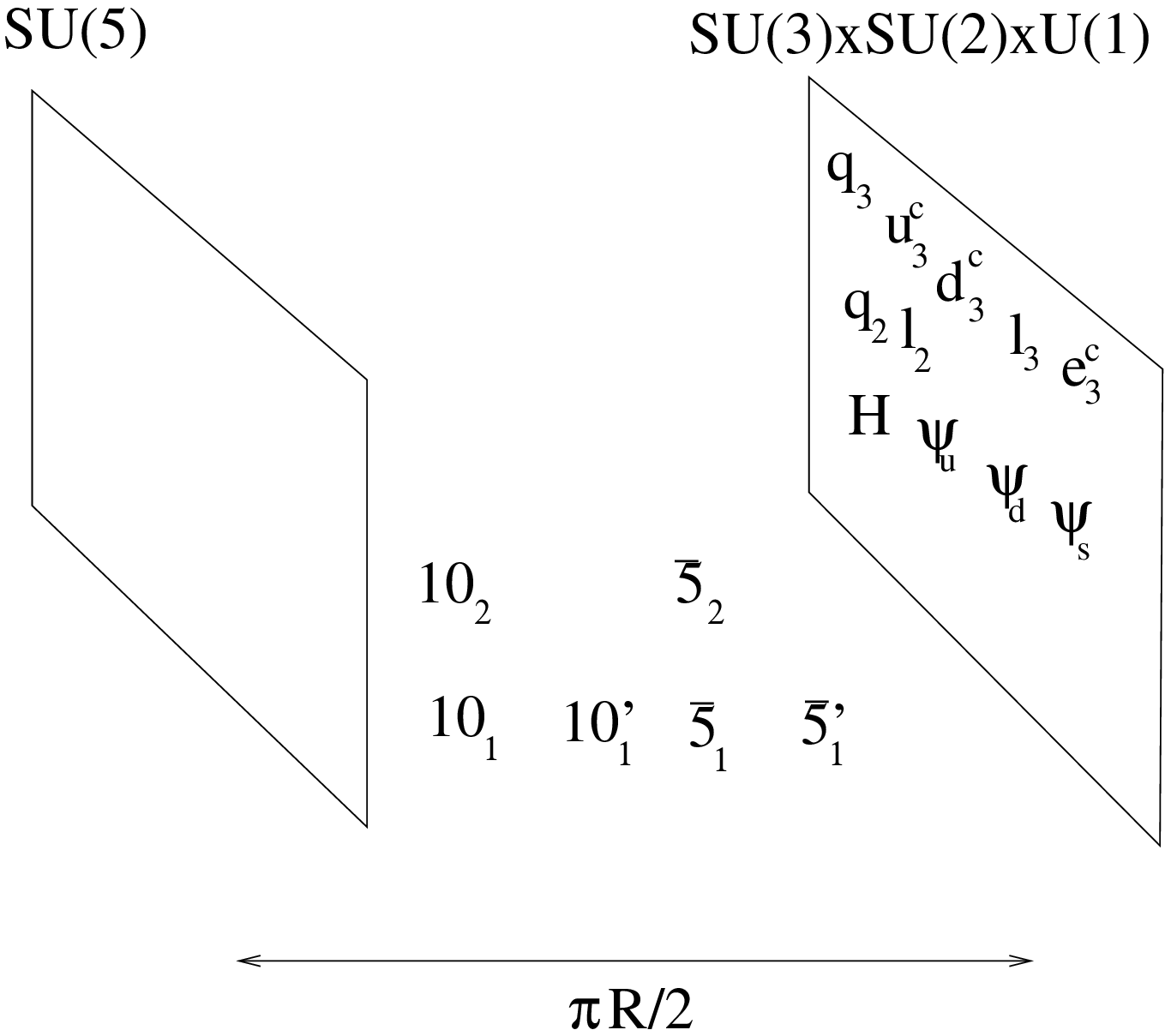}}
    \hspace{0.3in}
    \subfigure[Higgs in the bulk and third generation on the $SU(5)$ brane]
    {\label{bestmodel}\includegraphics[width=7 cm]{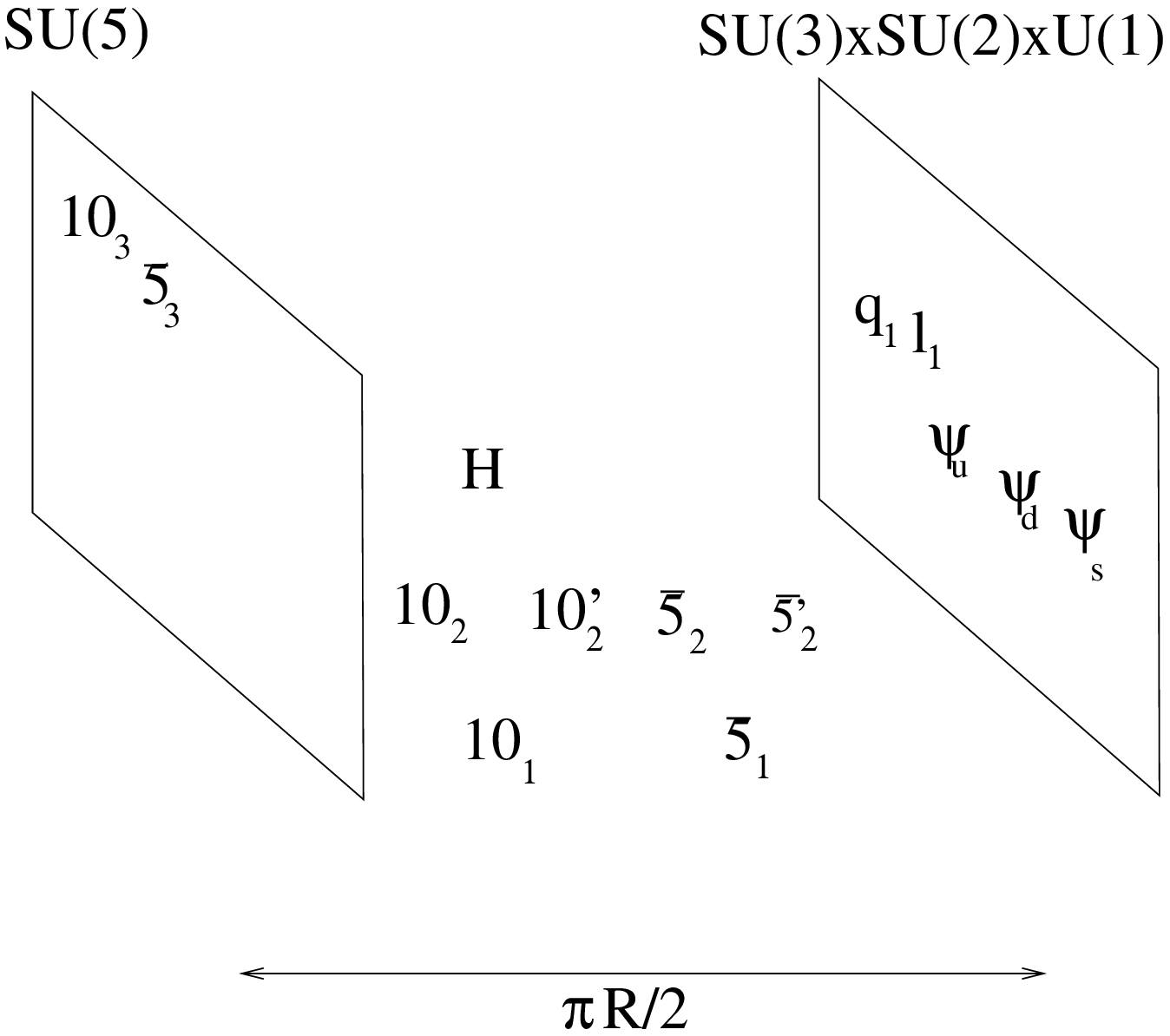}}
  \caption{\footnotesize Matter content of the two orbifold GUT models we propose.}
  \label{gutfig}
\end{figure}
\noindent In this model, which can be seen on the left-hand side of Figure \ref{gutfig}, we also need to put our
third generation and split second generation on the same brane in order for them to
interact with the higgs.  

On the right is
another model which capitalizes on every shred of evidence
we have about GUT physics:  we put the higgs in the bulk in this
case (recall that the orbifold naturally gives rise to
doublet-triplet splitting) so that we can switch the third generation to the
$SU(5)$-preserving brane and obtain $b$-$\tau$ unification
(see Figure \ref{btau}) without having analogous relationships
for the other two generations\footnote{As explained in
\cite{Hall:2002ci} the two yukawa couplings $\lambda_b$ and
$\lambda_{\tau}$ run differently only below the compactification
scale.  Because of locality, the fields living on the $SU(5)$
brane do not feel the $SU(5)$ breaking until energies below the
compactification scale; hence if they are unified at some high
energy they keep being unified until this scale.}.  We also need to flip the positions of the first and second generations if we want to keep the suppression of the mass of the first generation with respect to the second.

\begin{figure}[ht]
  \centering
    \includegraphics[width=10 cm]{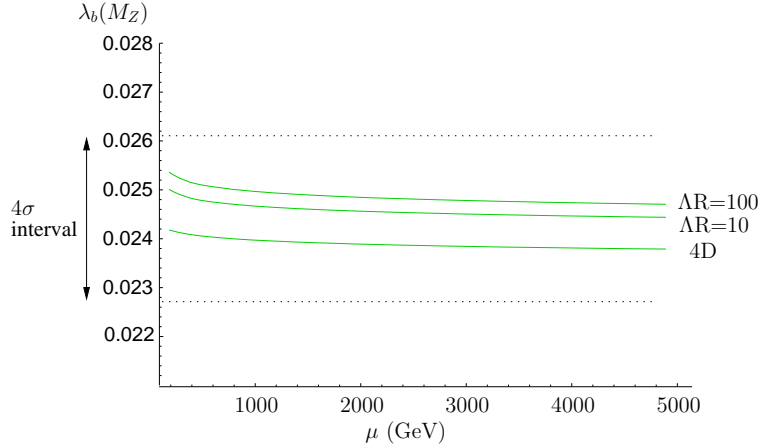}
  \caption{\footnotesize Low energy prediction for $\lambda_b(M_Z)$, as a function of the
higgsino mass $\mu$, for the model with the higgs in the bulk, for
$\Lambda R=10,100$ and for the 4D model.  4$\sigma$ interval taken from
\cite{Eidelman:2004wy}.}
  \label{btau}
\end{figure}

The low-energy values for $\alpha_3$ as a function of
$\mu$ in these two models can be seen in Figure \ref{alphastrongfig} for
different $\Lambda R$.  Note that unification can be acheived in the regime where $\Lambda R\gg 1$,
justifying our initial assumption that the brane kinetic terms could be neglected.  We see that although the dependence on $\mu$
is very slight, small $\mu$ seems to be preferred.  However we cannot
use this
observation to put a firm upper limit on $\mu$ because of the
uncertainties associated with ignoring the bare kinetic terms on the branes.

\begin{figure}[ht]
  \centering
    \includegraphics[width=12 cm]{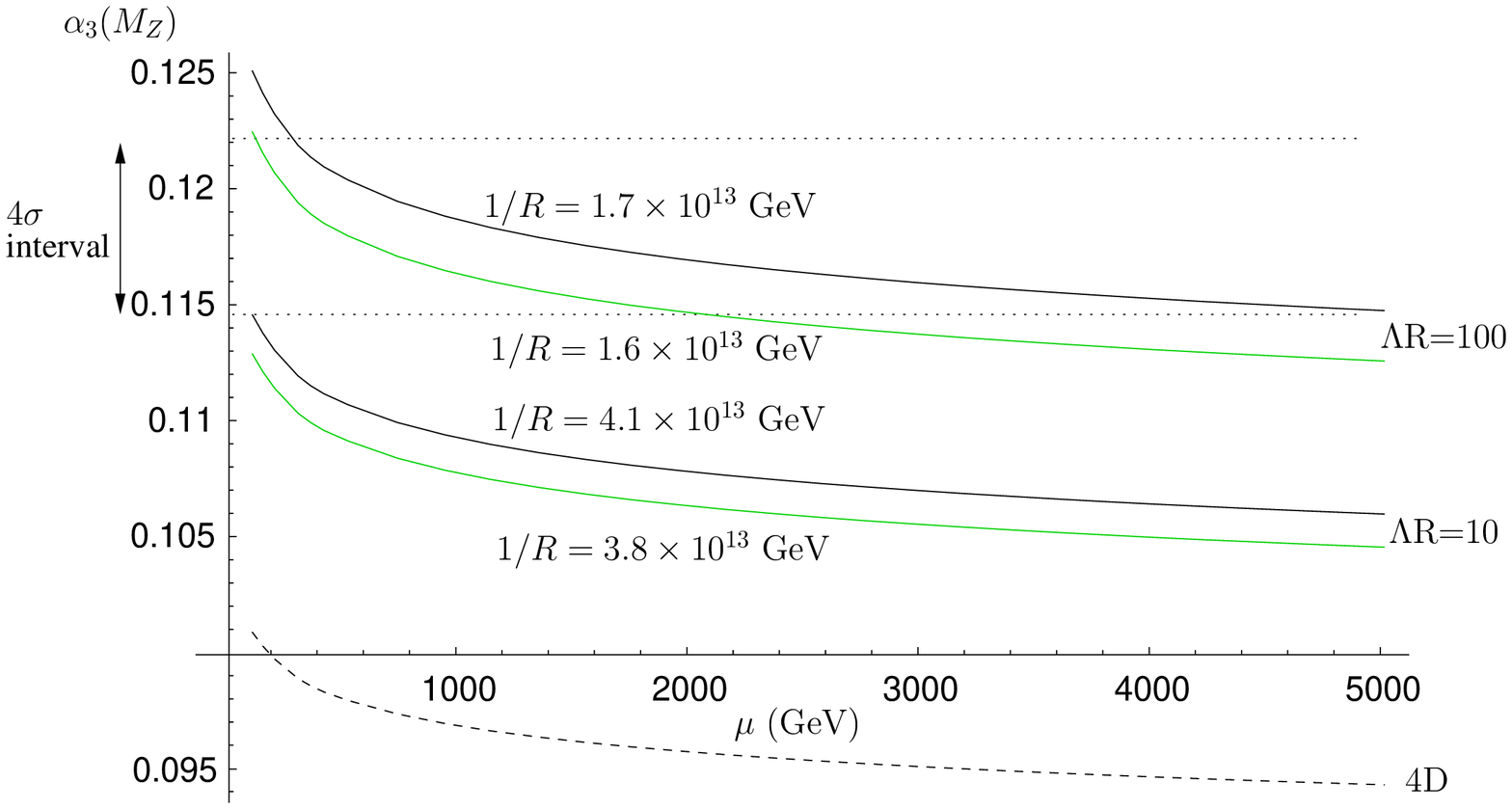}
  \caption{{\footnotesize Low energy prediction for $\alpha_3(M_z)$, as a function of the
higgsino mass $\mu$, for the model with the higgs on the brane (black line),
and for the model with the higgs in the bulk (green line), for
$\Lambda R=10,100$, and for the 4D case (dashed line). Some typical values of $1/R$ are shown.}}
  \label{alphastrongfig}
\end{figure}

The second configuration, Figure \ref{bestmodel}, also gives proton decay through the
mixing of the third generation with the first two.  From our
knowledge of the CKM matrix we infer that all mixing matrices
will be close to the unit matrix, proton decay will therefore be
suppressed by off-diagonal elements.  To minimize this suppression
it is best to have an anti-neutrino and a strange quark in the
final state. The proton decay rate for this process was computed
in \cite{Hebecker:2002rc} and is proportional to
$\left(\frac{g^2_4}{(1/R)^2}\right)^2\left(1-\frac{m^2_{\rm kaon}}
{m^2_{\rm proton}}\right)|(R^\dag_d)_{23}
(R^\dag_u)_{13}(L_d)_{31}|^2$ where $R_{d,u}$ and $L_d$ are the
rotation matrices of the right-handed down-type and up-type
quarks, and the left-handed down-type quarks, which are unknown.
We assume that
the 2-3 and the 1-3 mixing elements are 0.05 and 0.01
respectively, similar to the corresponding CKM matrix elements,
giving a proton lifetime of

\begin{equation}
\tau_p(p\rightarrow K^+\bar{\nu}_\tau) \simeq6.6\times 10^{38}{\rm
\ years \ }\times \left(\frac{1/R}{10^{14}{\rm
GeV}}\right)^4\simeq 4\times 10^{35}\ {\rm years}
\end{equation}

\noindent for $1/R=1.6\times 10^{13}$ GeV. This is above the current limit
from Super-Kamiokande of $1.9\times 10^{33}$ years at 90\% C.L.
\cite{Hayato:1999az,Suzuki:2001rb}, although there are
multi-megaton experiments in the planning stages that are expected
to reach a sensitivity of up to $6\times 10^{34}$ years
\cite{Suzuki:2001rb} .  Given our lack of information about
the mixing matrices involved\footnote{Experiments have only
constrained the particular combination that appears in the SM as the
CKM matrix.}, we see that there might be some possibility that
proton decay in this model will be seen in the not-too-distant
future.


\section{Conclusion \label{sec:conclusion}}

The identification of a TeV-scale weakly-interacting
particle as a good dark matter candidate, and the unification of the gauge
couplings are usually taken as indications of the
presence of low-energy SUSY.  However this might not necessarily be the case.

If we assume that the tuning of the higgs mass can be explained in
some other unnatural way, through environmental reasoning
for instance, then new possibilities open up for physics
beyond the SM.  In this paper we studied the minimal model
consistent with current experimental limits, that has
both a good thermal dark matter candidate and gauge coupling
unification.
To this end we added to the SM two higgsino-like particles and a
singlet, with a singlet majorana mass of $\lesssim 100$ TeV in order to split the two neutralinos and so
avoid direct detection constraints.  Making the singlet light allowed for a new region of dark matter
with mixed states as heavy as $\sim 2.2$ TeV, well beyond the
reach of the LHC and the generic expectation for a weakly
interacting particle.  Nevertheless we do have some handles on this
model: firstly via the 2-loop induced electron EDM contribution
which is just beyond present limits for CP angle of order 1, and
secondly by the spin-independent direct detection cross section,
both of which should be accessible at next-generation
experiments.

Turning to gauge coupling unification we saw that this was much
improved at two loops by the presence of the higgsinos.  A
full 4D GUT model is nevertheless excluded by the smallness of the GUT
scale $\sim 10^{14}$ GeV, which induces too fast proton decay. We
embedded the model in a 5D orbifold GUT in which the threshold
corrections were calculable and pushed $\alpha_3$ in the right
direction for unification (for a suitable matter configuration).
It is very gratifying that such a model can help explain the pattern in the fermion mass
hierarchy, give $b$-$\tau$ unification, and predict a rate
for proton decay that can be tested in the future.

\section*{Acknowledgments}
First and foremost we would like to thank Nima Arkani-Hamed for his constant guidance and insight.  Thanks also to Aaron Pierce for various helpful suggestions; and Paolo Creminelli, Alberto Nicolis, Philip Schuster, Jesse Thaler, and Natalia Toro for useful conversations. 

\appendix

\section*{Appendix}
\section{The neutralino mass matrix\label{app: neutralino mass matrix}}

We have a 3$\times$3 neutralino mass matrix in which the mixing terms
(see equation (\ref{mass_matrix})) are unrelated to gauge couplings
and are limited only by the requirement of perturbativity to the
cutoff.  It is possible to get a feel for the behavior of this matrix
by finding the approximate eigenvalues and eigenvectors in the limit
of equal and small off-diagonal terms ($\lambda v\ll M_1\pm\mu$).  The approximate eigenvalues and eigenvectors are shown in the table
below:
\\
\\
\begin{tabular}{|c|c|}
\hline
{$M^2$}&{Gaugino fraction}\\
\hline\hline
$M_1^2+4\lambda^2 v^2\frac{M_1}{M_1\pm\mu}$ & $1-2\frac{\lambda^2 v^2}{(M_1\pm\mu)^2}$\\
\hline
$\mu^2$ &   0\\
\hline
$\mu^2\pm 4\lambda^2 v^2\frac{\mu}{M_1\pm\mu}$ & $2\frac{\lambda^2 v^2}{(M_1\pm\mu)^2}$\\
\hline
\end{tabular}
\\
\\
\noindent for $\cos{\theta}=\pm 1$.

If $M_1<<\mu$ the first eigenstate will be the LSP.  In the opposite limit
the composition of the LSP is dependent upon the sign of $\cos(\theta)$.  For
$\cos(\theta)$ positive the second eigenstate, which is pure higgsino, is the LSP,
while for $\cos(\theta)$ negative, the mixed third eigenstate becomes the LSP.

We also see from the table that in the pure higgsino case, a splitting of order 100 keV (sufficient to evade the direct detection constraint) can be achieved with a singlet mass lighter than $10^9$ GeV, where the upper limit corresponds to O(1) yukawa couplings.

\section{Two-Loop Beta Functions for Gauge Couplings\label{app: beta functions}}

The two-loop RGE for the gauge couplings in our minimal model is

\begin{equation*}
(-2\pi)\frac{d}{dt}\alpha_i^{-1}=b^{MM}_i+\frac{1}{(4\pi)^2}\left[\sum^3_j 4\pi B^{MM}_{ij}\alpha_j-d_i \lambda_t^2-d'_i(\lambda_u^2+\lambda_d^2)\right]
\end{equation*}

\noindent with $\beta$-function coefficients

\begin{equation*}
b^{MM}=\left(\frac{9}{2},-\frac{15}{6},-7\right)\hspace{1cm}
B^{MM}= \left(\begin{array}{ccc}
\frac{104}{25} & \frac{18}{5} & \frac{44}{5} \\
\frac{6}{5} & 14 & 12 \\
\frac{11}{10} & \frac{9}{2} & -26
\end{array}
\right)\hspace{1cm}
d=\left(\frac{17}{10},\frac{3}{2},2\right)\hspace{1cm}
d'=\left(\frac{3}{20},\frac{1}{4},0\right)
\end{equation*}

The running of the yukawa couplings is the same as in \cite{Giudice:2004tc} but we will reproduce their RGEs here for convenience - we ignore all except the
top yukawa coupling (we found that our two new yukawas do not have a significant effect).

\begin{equation}
(4\pi)^2\lambda_t=\lambda_t\left[-3\sum^3_{i=1} 4\pi c_i\alpha_i+\frac{9}{2}\lambda_t^2+\frac{1}{2}(\lambda_u^2+\lambda_d^2)\right]\nonumber
\end{equation}

with $c=\left(\frac{17}{60},\frac{3}{4},\frac{8}{3}\right)$.

The two-loop coupled RGEs can be solved analytically if we
approximate the top yukawa coupling as a constant over the entire range of
integration (see \cite{Langacker:1992rq} for a study on the validity of this
approximation).  The solution is

\begin{equation*}
\alpha_i^{-1}(M)=\alpha_G^{-1}+\frac{1}{2\pi}b^{MM}\ln{\frac{\Lambda}{M}}+\frac{1}{4\pi}\sum^3_{j=1}\frac{B^{MM}_{ij}}{b^{MM}_j}\ln{\left(1+\frac{1}{4\pi}b^{MM}_j\alpha_G(\Lambda)\ln{\frac{\Lambda}{M}}\right)}-\frac{1}{32\pi^3}d_i \lambda_t^2\ln{\frac{\Lambda}{M}}
\end{equation*}

\end{document}